# Single-Layer MoS$_2$ Phototransistors**

Zongyou Yin[1,†], Hai Li[1,†], Hong Li,[2] Lin Jiang,[1] Yumeng Shi,[1] Yinghui Sun,[1] Gang Lu,[1] Qing Zhang,[2] Xiaodong Chen,[1] Hua Zhang[1,*]

[1]School of Materials Science and Engineering, Nanyang Technological University, 50 Nanyang Avenue, Singapore 639798.

[2]School of Electrical and Electronic Engineering, Nanyang Technological University, 50 Nanyang Avenue, Singapore 639798

*Corresponding Author.
Phone: +65-6790-5175. Fax: +65-6790-9081
E-mail: hzhang@ntu.edu.sg, hzhang166@yahoo.com
Homepage: http://www.ntu.edu.sg/home/hzhang/

[†] These authors contributed equally to this work.

**Abstract:**

A new phototransistor based on the mechanically-exfoliated single-layer $MoS_2$ nanosheet is fabricated and its light-induced electric properties are investigated in details. Photocurrent generated from the phototransistor is solely determined by the illuminated optical power at a constant drain or gate voltage. The switching behavior of photocurrent generation and annihilation can be completely finished within ca. 50 ms and it shows good stability. Especially, the single-layer $MoS_2$ phototransistor exhibits a better photoresponsivity as compared with the graphene-based device. The unique characteristics of incident-light control, prompt photoswitching and good photoresponsivity from the $MoS_2$ phototransistor pave an avenue to develop the single-layer semiconducting materials for multi-functional optoelectronic device applications in future.

As one of the most promising two-dimensional (2D) materials, graphene, one-atom-thick single layer of carbon atoms packed with honeycomb lattice, has shown exceptional physical, chemical, optical and mechanical properties.[1-13] Recently, other 2D materials have attracted increasing attention.[14-22] In particular, great interest has been focusing on the single-layer semiconducting materials, such as $MoS_2$, one of the transition-metal dichalcogenides, which exhibits the unique physical, optical and electrical properties correlated with its 2D ultra-thin atomic layer structure.[23-26] For example, the single-layer $MoS_2$ has a unique quantum luminescence efficiency[25,26] and exhibits a high channel mobility (~ 200 $cm^2V^{-1}s^{-1}$) and current ON/OFF ratio ($1\times10^8$) when it was used as the channel material in the field-effect transistor (FET).[24]

In this contribution, to the best of our knowledge, for the first time, we fabricate the single-layer $MoS_2$ based phototransistor and then study its light-induced electric properties in details. Based on our experimental results, the photocurrent ($I_{ph}$) generation efficiency is dependent on the magnitude of photo-generated charges under a constant drain voltage ($V_{ds}$) or gate voltage ($V_g$). Photocurrent generation and annihilation (called as photoswitching in the text) behavior is so prompt that the photocurrent can be completely switched between ON and OFF states within ca. 50 ms, which is fully controlled by the incident light and exhibits a quite stable characteristic. The photoresponsivity from the single-layer $MoS_2$ phototransistor can reach as high as 7.5 mA/W under illumination with a low optical power ($P_{light}$, 80 μW) and a medium gate voltage ($V_g$, 50 V). Importantly, the obtained photoresponsivity from the single-layer $MoS_2$ is better than that obtained from the single-layer graphene based FET (1 mA/W) which mainly arises from the rapid electron-hole recombination induced by the intrinsic property of zero-bandgap and fast carrier transfer of graphene.[27,28]

## RESULTS AND DISCUSSION

**Preparation and characterization of single-Layer MoS$_2$ based phototransistors.** In a typical experiment, the single-layer MoS$_2$ was deposited onto on a Si/SiO$_2$(300 nm) substrate by using the scotch-tape based mechanical exfoliation method.[29] The optical and AFM images of the obtained single-layer MoS$_2$ on Si/SiO$_2$ are shown in Figure 1A and B, respectively. The height of the single-layer MoS$_2$ measured by AFM is ca. 0.8 nm, which is consistent with the previous theoretical[30,31] and experimental[23-26] results. The photoluminescence (PL) of single-layer MoS$_2$ sheet was measured at room temperature using the 488 nm laser (Figure 1C). The dominated PL peak at 676 nm arises from the direct intraband recombination of the photo-generated electron-hole pairs in the single-layer MoS$_2$, and the weak shoulder peak at ca. 623 nm is attributed to the energy split of valence band spin-orbital coupling of MoS$_2$.[26] Raman spectrum was used to further confirm the single-layer MoS$_2$ (inset in Figure 1C). Two peaks at 384 and 400 cm$^{-1}$ are attributed to the in-plane $E^1_{2g}$ and out-of-plane $A_{1g}$ vibration of single-layer MoS$_2$, respectively.[20,29] Figure 1D shows the optical image of the single-layer MoS$_2$ FET device fabricated by photolithography, where the Ti/Au electrodes and Si were used as the source, drain and back gate, respectively.

The fabricated single-layer MoS$_2$ FET exhibits an obvious n-type semiconducting property (Figure 2), which is consistent with the previous reports.[23,24,29] Such n-type doping might come from the impurities, such as halogen (Cl or Br) atoms, which could replace S atoms in the natural MoS$_2$ crystals or exist as the interstitial atoms in the interlayer gap of MoS$_2$. This increases the total electron concentration of the host MoS$_2$ system resulting in an n-type doping for MoS$_2$ (see more detailed discussion in the Supporting Information). The field-effect mobility of this single-layer MoS$_2$ device can be estimated based on the equation,[24] $\mu = \frac{L}{W \times (\varepsilon_0 \varepsilon_r / d) \times V_{ds}} \times \frac{dI_{ds}}{dV_g}$, where the channel length $L$ is 2.1 µm, the channel width $W$ is 2.6 µm, $\varepsilon_0$ is 8.854×10$^{-12}$ Fm$^{-1}$, $\varepsilon_r$ for SiO$_2$ is 3.9, $d$ is the thickness of SiO$_2$ (300 nm). The calculated mobility of our device is ca. 0.11 cm$^2$V$^{-1}$s$^{-1}$, which is comparable with the previous results from the bottom-gate FET devices reported by other research groups.[23,24] However, the reported mobility is far lower than ~200 cm$^2$V$^{-1}$s obtained from the top-gate FET by deposition of the high-κ gate dielectric of HfO$_2$, used for the device mobility booster, on the top of MoS$_2$.[24] The reason is likely that the trap/impurity states exist at SiO$_2$ surface in the bottom gate FETs, and the scattering from these charged impurities degrades the device mobility.[24,32,33] Reduction of surface traps/impurities in the bottom gate dielectric is expected to improve the mobility of such single-layer MoS$_2$ based bottom-gate FET devices. The obtained drain current ON/OFF ratio of our device is ~10$^3$. Note that as compared to the mobility (ca. 0.03 cm$^2$V$^{-1}$s$^{-1}$) and current ON/OFF ratio (~10$^2$) from the single-layer MoS$_2$ FETs,[29] the enhanced mobility and ON/OFF ratio from the current single-layer MoS$_2$ device are attributed to the thermal annealing treatment after the device fabrication (see the Materials and Methods), which could remove the photoresist residue and improve the electrode contact.[24,34]

**Photocurrent dependence on the incident optical power.** The single-layer MoS$_2$ based phototransistor was investigated by exploring its output characteristics of photocurrent, including the generation, switching behavior and photoresponsivity, in various optical power and applied gate voltage. As shown in Figure 3A, when the excitation wavelength was above ~670 nm, the generated photocurrent, obtained after reduction of the dark drain current, was quite low. However when the wavelength decreased from 670 nm, the generated photocurrent increased obviously. This is reasonable since the photocurrent generation needs to match the basic condition, *i.e.* the incident photon energy must be greater than the energy gap ($E_g$) of single-layer $MoS_2$, which is around 1.83 eV, *i.e.* 676 nm in wavelength.[26] Only those incident photons with large energy ($h\nu > 1.83$ eV, *i.e.* wavelength < 676 nm) can excite electrons from the valence band (VB) to conduction band (CB) in the single-layer $MoS_2$, generating the photocurrent when the drain voltage is applied. On the other hand, when the incident photon energy is above the $E_g$ of $MoS_2$, The photocurrent is mainly dependent on the incident optical power and transition probability of photoexcited electrons (or photoelectrons). Photons with higher energy, *i.e.* shorter wavelength, will transfer more energy to electrons which have higher probability to successfully overcome the defect/impurity trapping of $MoS_2$, further overpass the Au:Ti/$MoS_2$ Schottky barrier and then flow to the external circuit to generate the photocurrent. In other words, the transition probability of photoelectrons, contributing to the photocurrent, is higher under the illumination of higher-energy photons. These high-energy photons can generate larger photocurrent as observed in Fig. 3A. However, increase of the photon energy leads to the decrease of its number under the constant optical power, which compromises the photocurrent increase. Therefore, the net increase rate of photocurrent becomes slow as the photon energy increases (Fig. 3A).

Under the constant excitation wavelength at 550 nm, even the optical power is less than 100 µW, the illumination on the single-layer MoS$_2$ can generate an obvious photocurrent which is power dependent. As shown in Figure 3B, when the optical power increases from 10 to 80 µW, the photocurrent increases gradually even the applied drain voltage is less than 1 V. In order to study the relationship between the output photocurrent ($I_{ph}$) and the incident optical power ($P_{light}$), the plot of $I_{ph}$ as function of $P_{light}$ is shown in Figure 3C, based on the results in Figure 3B. The $I_{ph}$ under the constant drain voltage is linearly proportional to $P_{light}$. Such relation has been described by the equation,[35] $I_{ph}=(q\mu_p pE)WD=CP_{light}$, where $q$ is the electronic charge, $\mu_p$ is the charge carrier mobility, $p$ is the charge carrier concentration, $E$ is the electrical field in the channel, $W$ is the gate width, $D$ is the depth of the absorption region, $C$ is the fitting parameter, and $P_{light}$ is the incident optical power. In such a single-layer MoS$_2$ based phototransistor, the linear relation of $I_{ph}$ *vs.* $P_{light}$ confirms that the photocurrent is solely determined by the amount of photo-generated carriers under illumination.

**Prompt photoswitching and good stability.** The photoswitching characteristic and stability of single-layer MoS$_2$ phototransistors were investigated at the room temperature in air. As shown in Figure 4A, the photocurrent as function of time was measured under the alternative dark and illumination conditions at different optical power and drain voltage. The switching behavior of drain current, *i.e.* the current ramps to a high value (ON state) under illumination and resumes to the low value (OFF state) under dark, was clearly observed. The generated photocurrent increases with the incident optical power and drain voltage. For example, when the optical power changes from 30 to 80 µW at constant drain voltage of 1 V, as expected, the photocurrent increases from 1.0 to 3.1 nA.

Note that the photocurrent further increases from 3.1 to 77.5 nA when the drain voltage increases from 1 to 7 V at constant optical power of 80 µW (Figure 4A). Such drain voltage dependent photocurrent generation indicates that some photo-generated charge carriers cannot be converted to the photocurrent when the applied drain voltage is low. This is reasonable since larger drain voltage can better drive photo-generated charges to electrode, or suppresses photo-generated charges from the recombination.

In order to study the photoswitching behavior of single-layer $MoS_2$ phototransistor, the change of photocurrent was recorded in a short time scale (Figure 4B). The observed switching duration for the current rise (from OFF to ON) or decay (from ON to OFF) process is only ca. 50 ms. In particular, such prompt photoswitching behavior can be obtained at different $V_{ds}$ or $P_{light}$. Moreover, the stability of this switching behavior was demonstrated by applying multiple illumination on the device for ~50 s. As shown in Figure 4C, the ON-OFF switching behavior can be well retained even after 20-cycle repeats. However, the response rate of photocurrent in our single-layer $MoS_2$ is still lower than that from graphene (tens of picoseconds), as the carrier transport in graphene is ballistic and very fast.[27,28]

**Good photoresponsivity tailored by gate voltage.** The photocurrent generation efficiency can be further enhanced by introduction of the gate voltage. In a typical experiment, if the gate voltage varies from -30 to 50 V, while $P_{light}$ and $V_{ds}$ are kept at 80 µW and 1 V, respectively, the recorded drain current under illumination is higher than that under dark (Figure 5A). Moreover, the prompt photocurrent ON/OFF switching behavior is still well maintained in the range of applied gate voltage, which could be seen from the similar prompt photocurrent switching results under the different gate voltage as shown in Figure 5B. All these results further confirm the good stability of our single-layer MoS$_2$ phototransistor and also demonstrate that the photocurrent can be tailored with the incident optical power, drain or gain voltage.

Photoresponsivity, a critical parameter to evaluate the performance of a phototransistor, is defined as photoresponsivity = $I_{ph}/P_{light}$, where $I_{ph}$ is the photocurrent generated and $P_{light}$ is the total incident optical power on the MoS$_2$.[27] Based on the photocurrent generated under different gate voltage, the calculated photoresponsivity as function of the gate voltage is plotted in Figure 5C. Under the zero gate voltage, the photoresponsivity is ~ 0.42 mA/W with $P_{light}$ and $V_{ds}$ set at 80 µW and 1 V, respectively. But it reaches as high as 7.5 mA/W at the gate voltage of ~50 V, proving that the back gate plays an important role in tailoring the photocurrent in the n-type single-layer MoS$_2$ phototransistor. Admittedly, the photoresponsivity from our single layer MoS$_2$ phototransistor is low as compared to the reported phototransistors based on ZnO nanowires (1.29×10$^4$ A/W) or vertical Si nanowire arrays (~ 10$^5$ A/W),[36,37] but it is higher than that in the 2D graphene based devices, ~1 mA/W at gate voltage of 60 V.[27,28] The lower photoresponsivity from graphene, as compared with our single-layer MoS$_2$, is probably due to the intrinsic zero bandgap, fast carrier transport and a short photocarrier life time in graphene which may trigger a fast recombination of the photo-generated carriers. The higher photoresponsivity from the one-dimensional ZnO or Si nanowire based phototransistors, as compared to single layer MoS$_2$ or graphene based devices, might be due to the material dimensions or other factors, which needs the further study.

The gate voltage dependent photoresponsivity in our single-layer MoS$_2$ photransistor is attributed to the n-type doping of MoS$_2$. As schematically illustrated in Figure 5D, before contact, Au and MoS$_2$ have different Fermi level ($E_F$, Figure 5D-I). The equilibrium state will be reached upon the contact between Au and MoS$_2$, thus forming a new quasi Fermi level ($E_{QF}$, Figure 5D-II). Note that, here, the gate voltage has not been applied and the Ti bonding layer is not considered in terms of Fermi level since its thickness is only ca. 3 nm, much thinner than the Au electrode layer (50 nm). However, if the negative gate voltage is applied, $E_{QF}$ moves from CB to VB of MoS$_2$, resulting in a larger barrier between the CB of MoS$_2$ and the $E_{QF}$ of Au electrode (Figure 5D-III) as compared to the state at zero gate voltage (Figure 5D-II). Therefore, the photo-generated charges difficultly drift to the external circuit, resulting in a low photocurrent generated. On the contrary, if the applied gate is positive, $E_{QF}$ approaches closer to the CB of MoS$_2$, forming a smaller barrier between the CB of MoS$_2$ and the $E_{QF}$ of Au electrode (Figure 5D-IV) as compared with the state under zero gate voltage (Figure 5D-II), leading to the photo-generated charges which drift efficiently to the external circuit to produce a high photocurrent.

**CONCLUSIONS**

In summary, for the first time, we fabricate the mechanically-exfoliated single-layer MoS$_2$ based phototransistor and investigate its electric characteristics in details. The photocurrent generation solely depends on the illuminating optical power at a constant drain or gate voltage. Photocurrent generation and annihilation can be switched within ca. 50 ms. Such prompt photoswitching behavior, controllable by the incident light, exhibits stable characteristic. Importantly, the single-layer MoS$_2$ phototransistor gives a better photoresponsivity than does the graphene-based device. Overall, the unique characteristics, such as incident-light control, prompt photoswitching and good photoresponsivity, of the single-layer $MoS_2$ phototransistor encourage one to fabricate the optoelectronic device based on the single atomic layer semiconductors by a facile and low-cost mechanic exfoliation method. This work opens an avenue to develop the single-layer semiconducting materials for the future functional device applications in the switches, memories, signal-amplifiers, light-related sensors, *etc*.

**MATERIALS and METHODS**

**Deposition of $MoS_2$ on $Si/SiO_2$ substrate.** Single-layer $MoS_2$ sheet was peeled from bulk $MoS_2$ (429MM-AB, SPI Molybdenum Disulfide, natural single crystals from Canada, CAS#: 1317-33-5, RTECS No.: QA4697000) and deposited onto a cleaned $Si/SiO_2$(300 nm) substrate using scotch tape-based exfoliation.[29] The optical microscope (Eclipse LV100, Nikon) was used to locate the target $MoS_2$ sheets. AFM (Dimension 3100 with Nanoscope IIIa controller, Veeco), working under the tapping mode in air, was used to measure the thickness of $MoS_2$ sheets so as to confirm their layer numbers.

**Photoluminescence and Raman spectroscopy.** The single-layer $MoS_2$ were characterized by Photoluminescence (PL) and Raman spectroscopy at room temperature (both measured on a WITec CRM200 confocal Raman microscopy system) at the excitation line of 488 nm and an air cooling charge coupled device (CCD) as the detector (WITec Instruments Corp, Germany). The Raman band of Si at 520 $cm^{-1}$ was used as a reference to calibrate the spectrometer.

**Fabrication and characterization of single-layer $MoS_2$ phototransistors.** The phototransistor has the similar configuration of the field-effect transistor (FET) while the incident illumination light source was used to tailor the photocurrent. The light source was set up based on the monochromator which has an adjustable wavelength of 400-1000 nm and optical power of 10-80 µW. The spot size of the incident light is ~ 10×10 $\mu m^2$, which covers the full $MoS_2$ channel and both electrodes. Note that the dominant photocurrent in the circuit is from $MoS_2$ since the same work function material Au:Ti is used as the source and drain electrodes. The electrode patterns for the source and drain of device were defined by the conventional photolithography and the lift-off process. The 3-

nm-Ti/50-nm-Au source and drain electrodes were deposited by electron beam evaporation. After removing the photoresist by acetone, the fabricated phototransistor were annealed for 2 h in a vacuum tube furnace with 100 sccm Ar:$H_2$ (v/v=9:1) flow at 200 °C to remove the photoresist residue and improve the contact for devices.[24,34] Finally, the electrical properties of the single-layer $MoS_2$ phototransistor were test at room temperature in air.

**Acknowledgement.** This work was supported by AcRF Tier 2 (ARC 10/10, No. MOE2010-T2-1-060) from MOE, CREATE program (Nanomaterials for Energy and Water Management) from NRF, and New Initiative Fund FY 2010 (M58120031) from NTU in Singapore.

**FIGURE CAPTIONS**

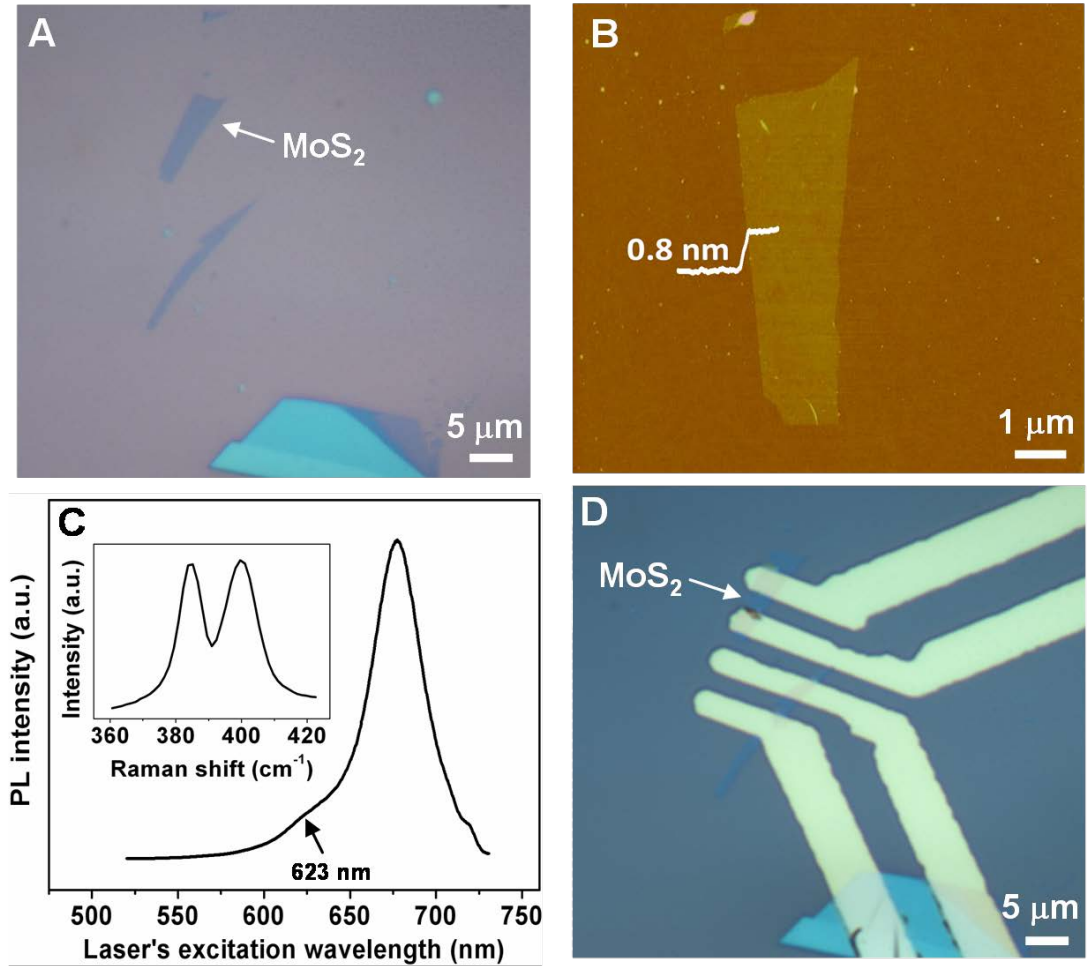

**Figure 1.** A) Optical and B) AFM images of single-layer $MoS_2$. C) PL and Raman (inset) spectra of single-layer $MoS_2$. D) Optical image of single-layer $MoS_2$ FET device.

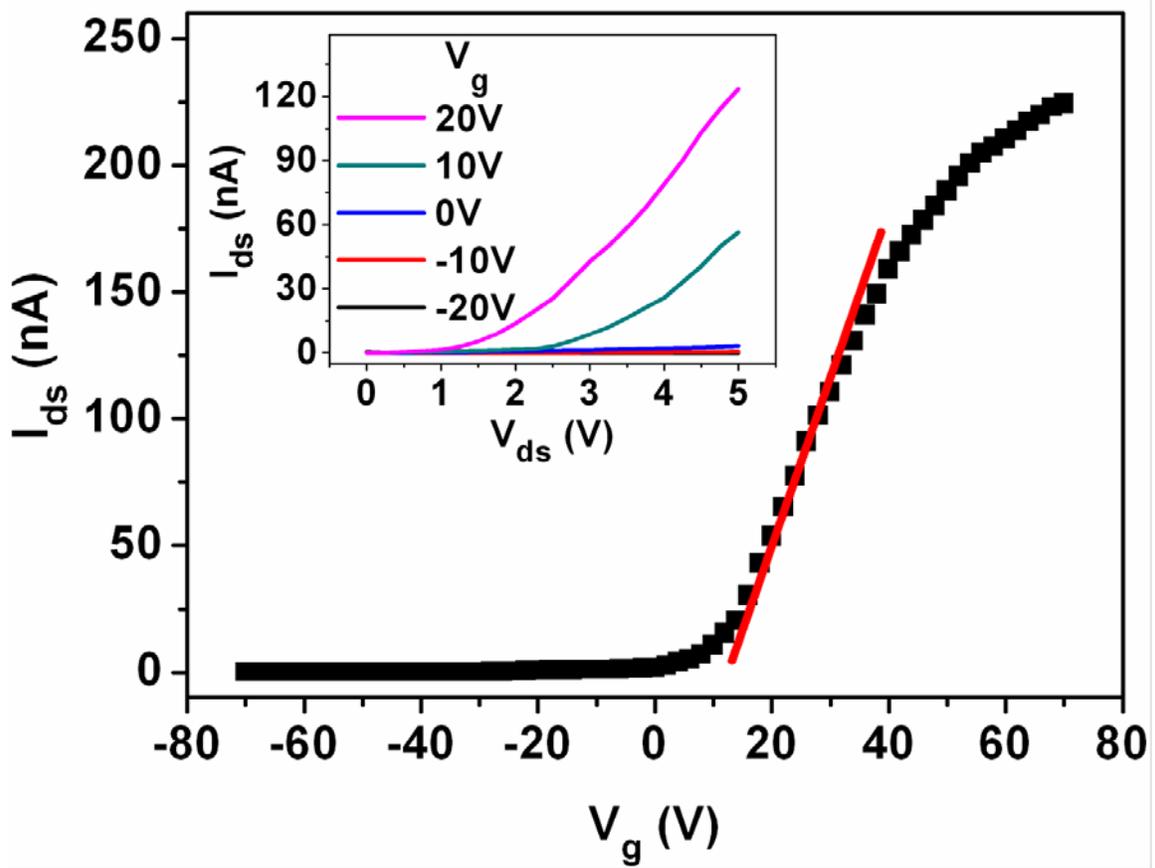

**Figure 2.** Room-temperature electrical characteristic of single-layer MoS$_2$ FET at drain voltage of 3 V. The red line is used to calculate the channel mobility. Inset: Plot of drain current *vs* voltage at different gate voltage (from -20 to 20 V).

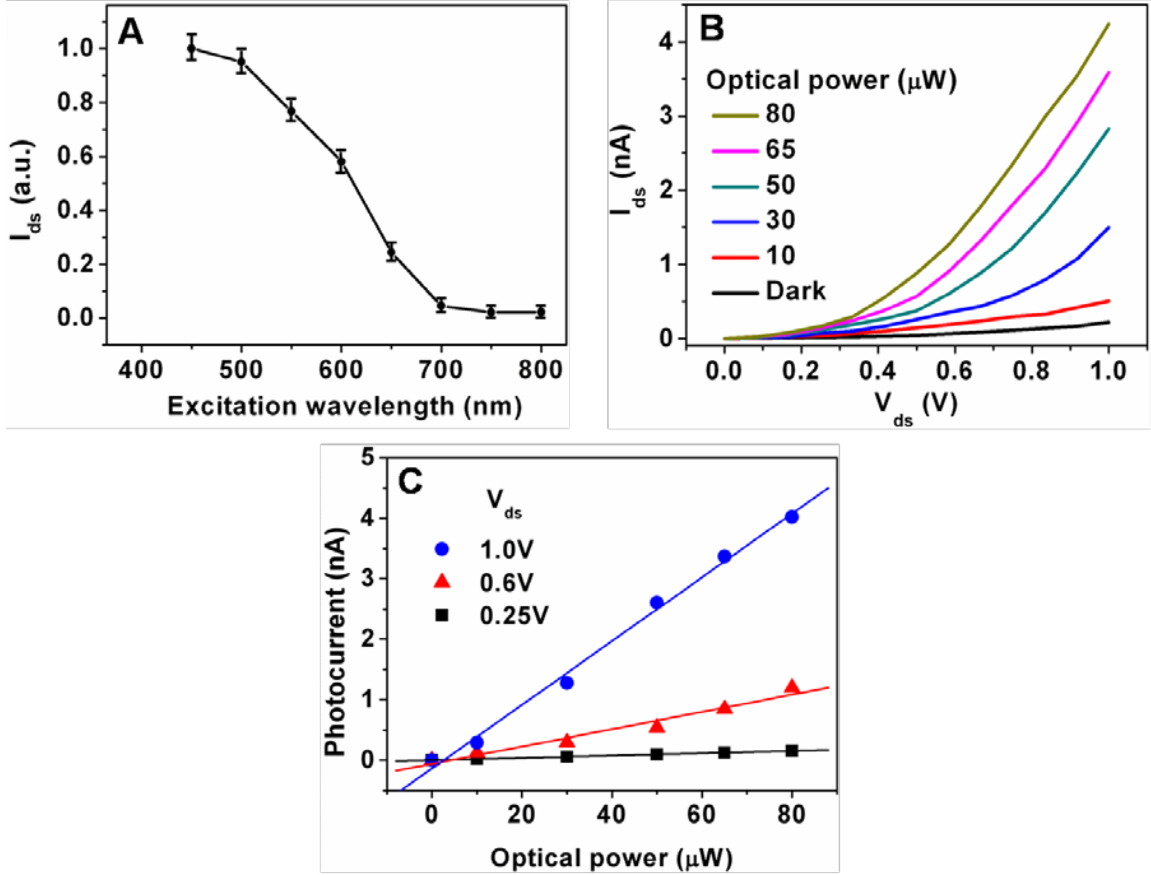

**Figure 3.** A) Drain current ($I_{ds}$) as function of excitation wavelength of the illumination source at a constant optical power of 80 μW. The $I_{ds}$ values were obtained by measuring three single-layer MoS$_2$ phototransistors. B) Typical output characteristics of phototransistor at different illuminating optical power (10 to 80 μW) at $V_g = 0$ V. C) Dependence of photocurrent on optical power at different $V_{ds}$ (0.25, 0.6 and 1.0 V). The linear curves are fitting results.

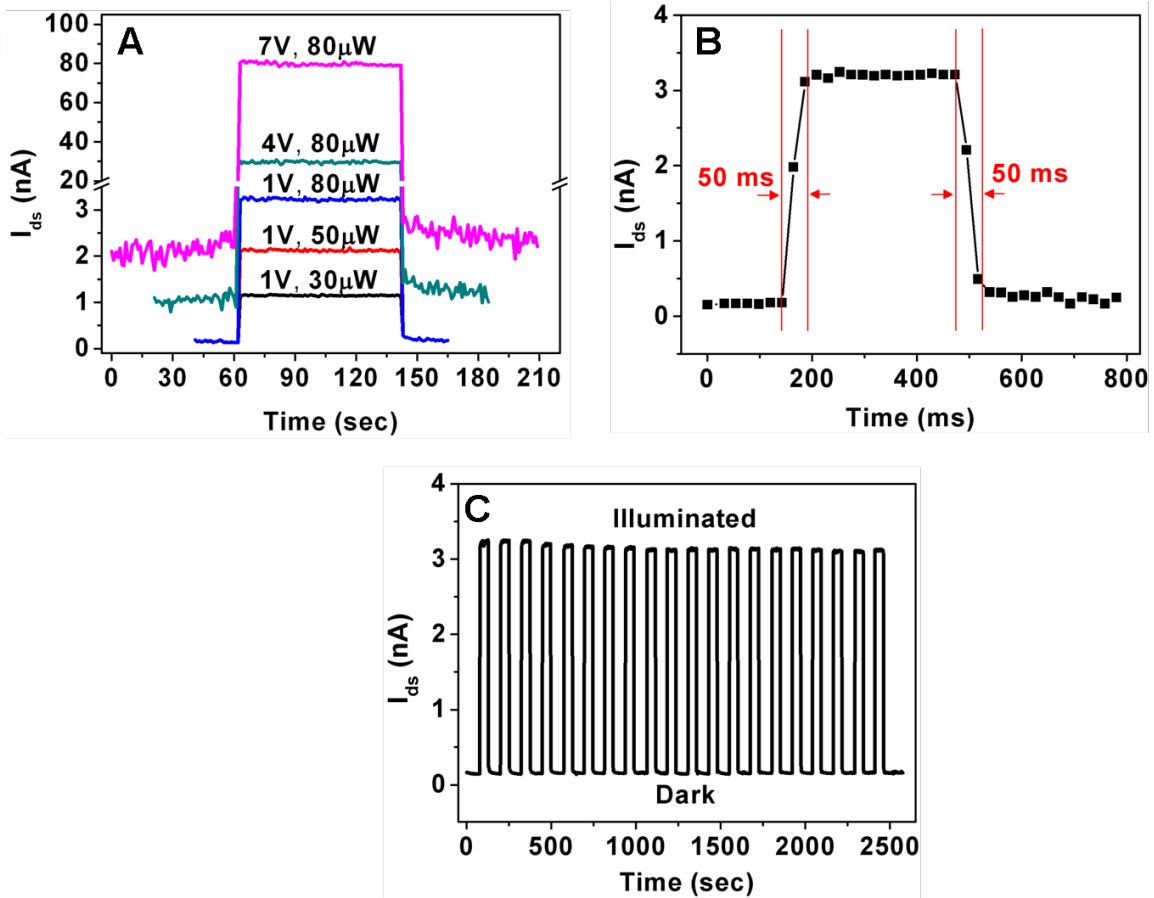

**Figure 4.** A) Photoswitching characteristics of single-layer MoS$_2$ phototransistor at different optical power and drain voltage. B) Photoswitching rate and C) stability test of photoswitching behavior of single-layer MoS$_2$ phototransistor at $V_{ds}$ = 1 V, $P_{light}$ = 80 μW.

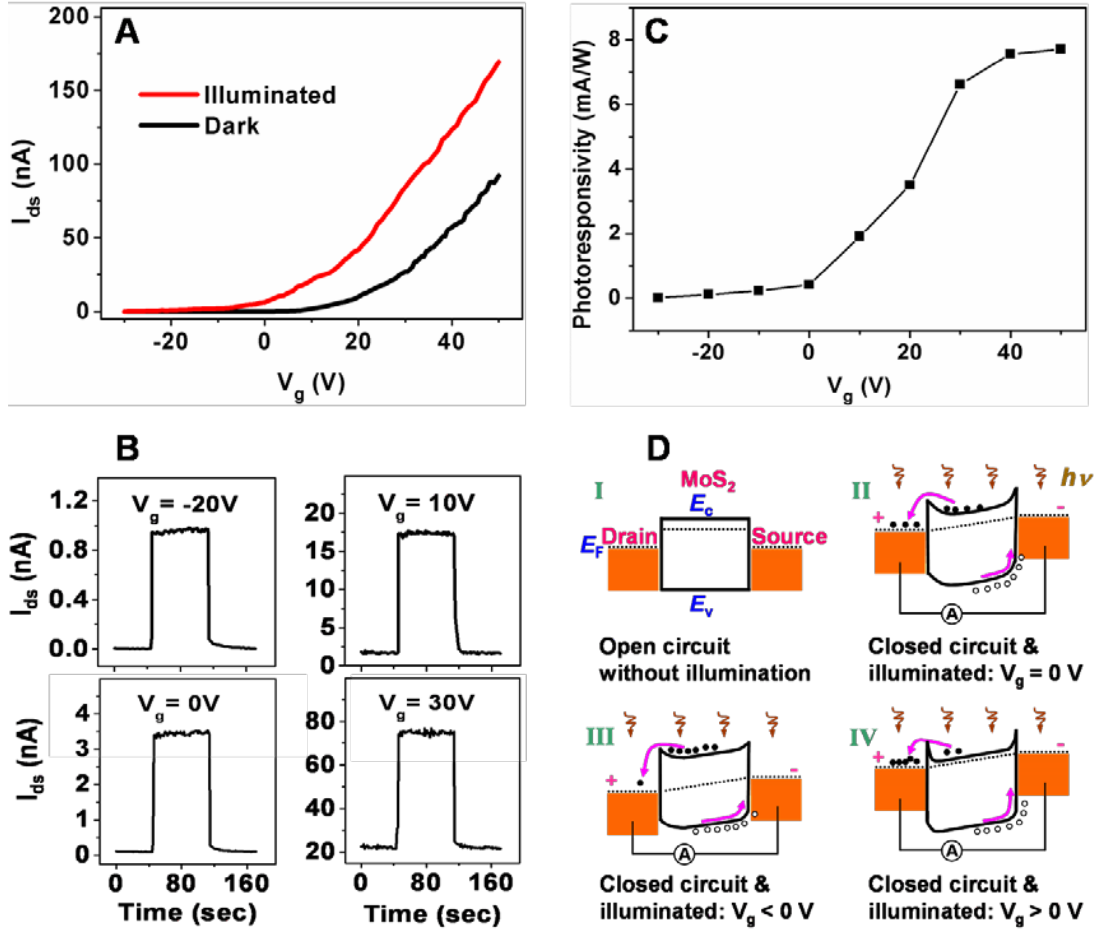

**Figure 5.** A) Typical output characteristics of single-layer MoS$_2$ phototransistor at gate voltage varied from -30 to 50 V ($P_{light}$ = 80 μW). B) Similar prompt photoswitching behavior at different gate voltage ($V_{ds}$ = 1 V, $P_{light}$ = 80 μW). C) Dependence of photoresponsivity on the gate voltage ($V_{ds}$ = 1 V, $P_{light}$ = 80 μW). D) Diagram of single-layer MoS$_2$ phototransistor circuit: Diagram I represents the initial state of device under open circuit without illumination. Diagrams II, III and IV represent the device under short circuit (source-drain voltage $V_{ds}$ = 1 V) and illumination ($P_{light}$ = 80 μW) while different gate voltages ($V_g$) were applied. The full dots at conduction band energy level ($E_c$) and the empty dots at valence band energy level ($E_v$) in MoS$_2$ represent the photoexcited electrons and holes, respectively. The conduction band energy level of $MoS_2$ is set at ~4.5 eV.[38] The valence band energy level of single-layer $MoS_2$ is ~6.3 eV because of its energy gap ~1.8 eV. The Fermi level of $MoS_2$ is set at 4.7 eV. The Fermi level of Au is 5.1 eV.

**SYNOPSIS TOC**

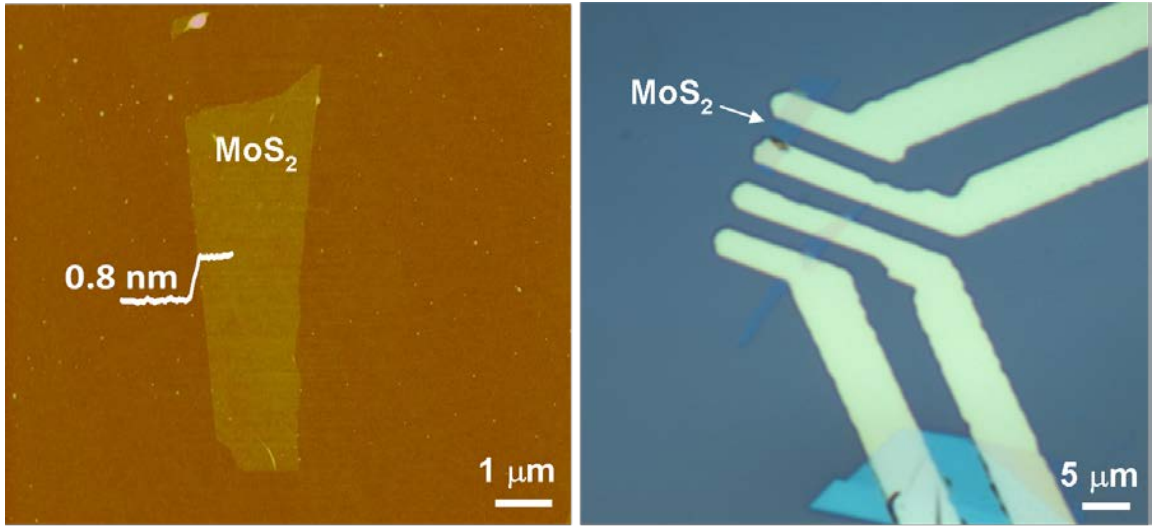

# Supporting Information

# Single-Layer MoS$_2$ Phototransistors

Zongyou Yin,[1,†] Hai Li,[1,†] Hong Li,[2] Lin Jiang,[1] Yumeng Shi,[1] Yinghui Sun,[1] Gang Lu,[1] Qing Zhang,[2] Xiaodong Chen,[1] Hua Zhang[1,*]

[1]School of Materials Science and Engineering, Nanyang Technological University, 50 Nanyang Avenue, Singapore 639798.

[2]School of Electrical and Electronic Engineering, Nanyang Technological University, 50 Nanyang Avenue, Singapore 639798.

*Corresponding Author.
Phone: +65-6790-5175. Fax: +65-6790-9081
E-mail: hzhang@ntu.edu.sg, hzhang166@yahoo.com
Homepage: http://www.ntu.edu.sg/home/hzhang/

† These authors contributed equally to this work.

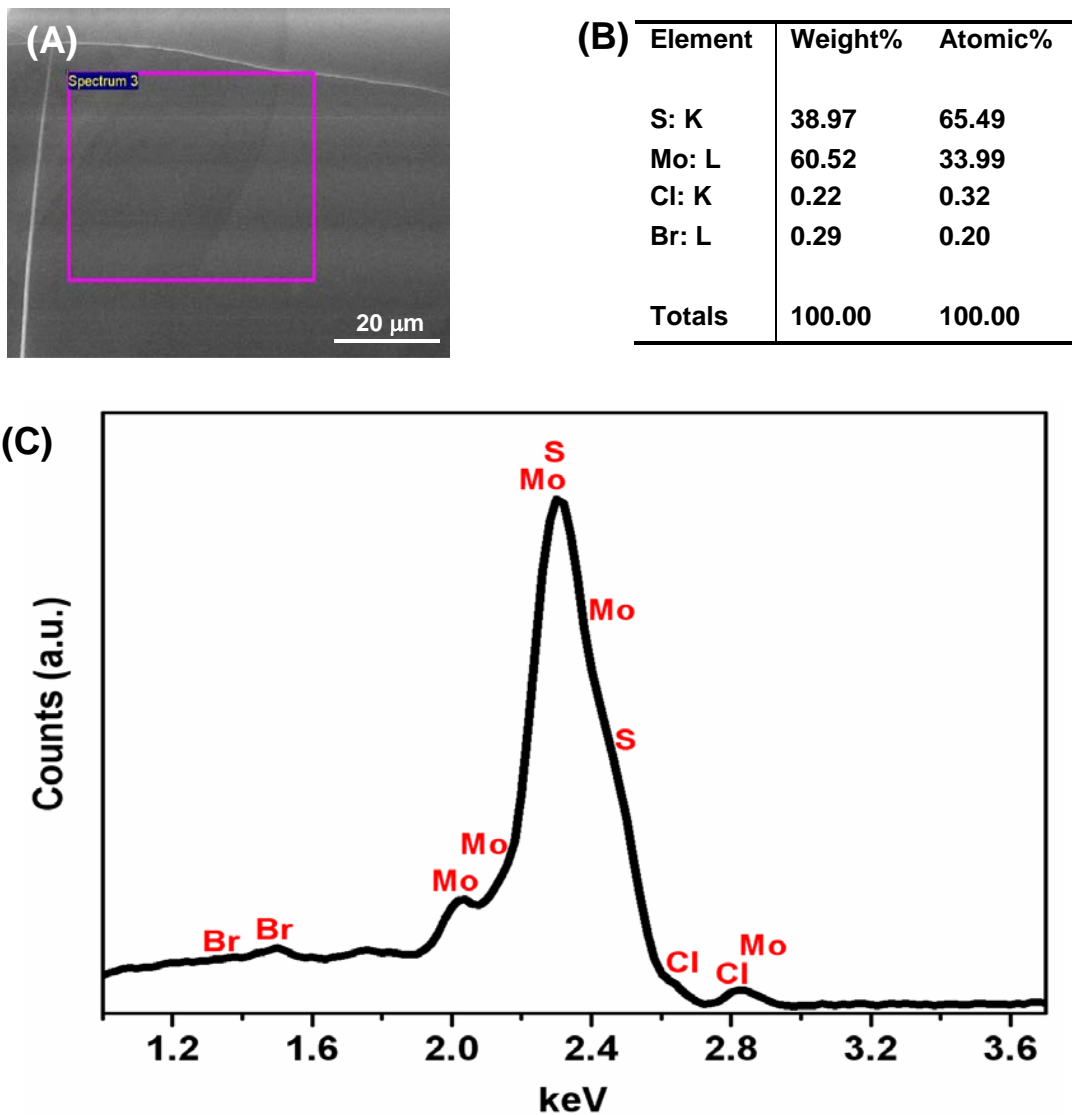

**Figure S1.** (A) SEM image of the natural MoS$_2$ crystal. (B) The weight and atomic ratio of MoS$_2$ crystal obtained from the energy dispersive X-ray (EDX) spectroscopy. (C) EDX spectrum with logarithmic scale on y axis labeled with S, Mo, Cl and Br elements.

The energy dispersive X-ray (EDX) spectroscopy was used to measure the MoS$_2$ crystal used in this work. The JEOL JSM-7600F field-emission scanning electron microanalyzer with an accelerating voltage of 20 kV was used. From the EDX results shown in Figure S1, the existence of both Cl and Br elements with quite low atomic ratio (~0.2-0.3%) were found. Note that such results only tell us the existence of Cl and Br elements, and the quantity of halogen elements is not accurate since EDX is a semi-quantitative analysis method. Based on these results, we might attribute the n-type impurities/donors of our MoS$_2$ flakes to the halogen atoms, such as Cl and Br atoms. These halogen atoms may work as the substitutional or interstitial doping for MoS$_2$.

In the substitutional doping mode, the halogen (Br or Cl) atoms may partially substitute S atoms during the formation/growth of the MoS$_2$ crystal in nature, which can work as the donors for MoS$_2$, since the doped halogen atoms, with one more electron than the S atoms, have increased the total electron concentration of MoS$_2$, resulting in an n-type doping for MoS$_2$. Doping Cl or Br atoms into MoS$_2$ in the substitutional doping mode to introduce the n-type MoS$_2$ has already been experimentally realized in the MoS$_2$ crystal growth with the aid of halogen vapor transport.[1,2] In the interstitial doping, the halogen (Br or Cl) atoms maybe present as the interstitial atoms in the interlayer gap of MoS$_2$, also resulting in the n-type doping for MoS$_2$. The interstitial doping of foreign atoms into layered transition metal dichalcogenides including MoS$_2$ has also been reported.[3-5]

---

[1] El-Mahalawy, S. H.; Evans B. L. Temperature Dependence of the Electrical Conductivity and Hall Coefficient in 2H-MoS$_2$, MoSe$_2$, WSe$_2$ and MoTe$_2$. *Phys. Stat. Sol. (b),* **1977**, *79*, 713-722.